\documentclass[english,preprint]{revtex4-1}
\usepackage[T1]{fontenc}
\usepackage[latin9]{inputenc}
\usepackage{mathrsfs}
\usepackage{amsmath}
\usepackage{amssymb}
\usepackage{graphicx}
\usepackage{esint}

\makeatletter

\newcommand{\mathcircumflex}[0]{\mbox{\^{}}}

\makeatother

\usepackage{babel}
\begin{document}

\preprint{BROWN-HET-1679}

\title{de Sitter gravity/Euclidean conformal gravity correspondence}

\author{Atreya Chatterjee}

\affiliation{Physics Department, Brown University, Providence, RI, 02912, USA}

\author{David A. Lowe}

\affiliation{Physics Department, Brown University, Providence, RI, 02912, USA}
\begin{abstract}
The holographic dual of a gravitational theory around the de Sitter
background is argued to be a Euclidean conformal gravity theory in
one fewer dimensions. The measure for the holographic theory naturally
includes a sum over topologies as well as conformal structures.
\end{abstract}
\maketitle

\section{Introduction}

There has been much success in describing gravity in Anti-de-Sitter
spacetime using a holographic description on the boundary at infinity.
In the holographic description, a conformal field theory (CFT) lives
on the boundary. This marks a major step in quantizing gravity in
asymptotically anti-de-Sitter spacetime. However according to experimental
observation, our universe has positive cosmological constant. Thus
it is interesting to consider a holographic description of gravity
in the de-Sitter spacetime.

One way of approaching this problem is to analytically continue the
AdS/CFT correspondence to dS/CFT correspondence \cite{Strominger:2001pn}.
There are many successes in this approach but there are many conceptual
difficulties as well. In this paper we will try to clarify some of
these difficulties. In the paper \cite{Chatterjee:2015pha}, we constructed
a mapping between bulk field operators and boundary operators. As
we saw in that paper, the boundary CFT has operators which violate
cluster decomposition. Cluster decomposition is one of the basic assumptions
of any interacting quantum field theory \cite{Osterwalder:1973dx,Osterwalder:1974tc}. 

To set the stage for understanding the problem in de Sitter spacetime,
be begin by considering the well-understood problem in anti de Sitter
spacetime. In that case, there is a positive energy theorem \cite{Abbott:1981ff,Breitenlohner:1982jf}
and the unitary representations of the conformal group $SO(d-1,2)$
that appear are lowest weight. Moreover the boundary conditions on
conformal infinity $\mathscr{I}$ that preserve conformal flatness
are compatible with the unitarity bound of \cite{Breitenlohner:1982jf}.
In particular, with these boundary conditions, one obtains a complete
set of modes for fluctuations around the anti-de Sitter background.

For de Sitter spacetime there is no global positive energy theorem
\cite{Abbott:1981ff} and the unitary representations of the conformal
group $SO(d,1)$ corresponding to ordinary massive and massless fields
are neither highest nor lowest weight, but are rather the principal
series and the complementary series, which are unbounded. This leads
to the problem of cluster decomposition violation in the boundary
theory, noted in \cite{Chatterjee:2015pha}. In the case of de Sitter,
a complete set of modes (for the graviton) leads to configurations
with a nontrivial conformal class at conformal infinity $\mathcal{\mathscr{I}}$.
Thus one cannot impose boundary conditions to maintain conformal flatness,
without truncating the linearized spectrum of the theory \cite{Ashtekar:2014zfa,Ashtekar:2015lla}.
Therefore to describe a quantum theory with the full set of modes
in a de Sitter background, the holographic description must accommodate
a path integral over boundary metrics. The boundary theory will be
invariant under the asymptotic symmetry group of the de Sitter spacetime
that preserves this more general set of asymptotic boundary conditions.
In this case, the asymptotic symmetry group is not just the conformal
group, corresponding to isometries of de Sitter, but is rather the
full group of diffeomorphisms of $\mathcal{\mathscr{I}}$. This leads
us to conjecture the holographic dual will be a theory of conformal
gravity theory living at $\mathcal{\mathscr{I}}$.

At first sight, this might seem a step backward, since theories of
conformal gravity seem difficult to quantize \cite{Fradkin:1985am}.
Nevertheless, there are examples where progress has been made. For
three dimensional pure conformal gravity, a Chern-Simons gauge formulation
is available \cite{Horne:1988jf}. For conformal gravity arising in
string theory, a twistor string formulation has been found \cite{Berkovits:2004jj}.
So there is hope that the rather different conformal gravity theories
considered here can be successfully quantized.

Having found a path integral over conformal classes of metric on $\mathcal{\mathscr{I}}$
is needed to provide a holographic description of gravity in de Sitter,
it is then natural to ask whether one must include a sum over topologies
of $\mathcal{\mathscr{I}}$ as well. In the case of anti-de Sitter,
this question was addressed in \cite{Witten:1999xp}. There it was
found that if $\mathcal{\mathscr{I}}$ has positive curvature, it
must be connected and cannot contain nontrivial topology, such as
wormholes. This result is important for the basic consistency of AdS/CFT.

Some related questions have been considered in the context of dS/CFT
in \cite{Andersson:2002nr,Galloway:2004bk}. However there it quickly
becomes clear that ordinary matter will lead to nontrivial topology
for $\mathcal{\mathscr{I}}$ in four-dimensional de Sitter since a
black hole already changes the topology from $S^{3}$ for empty de
Sitter to $S^{2}\times R$ for a black hole. Recall in AdS, the topology
of $\mathcal{\mathscr{I}}$ remains $S^{2}\times R$ for empty AdS,
or the AdS Schwarzschild black hole. 

One can gain a more detailed understanding of this topology change
in the case of three-dimensional de Sitter. As an example, we consider
the solution for multi-black holes in three-dimensional de Sitter
spacetime \cite{Deser:1983nh}. We show $\mathcal{\mathscr{I}}$ can
be mapped from a multi-sheeted sphere to a single cover with punctures.
The resulting holographic dual is a theory of two-dimensional gravity,
identical to a worldsheet string theory. At least in this example,
there is a natural moduli space corresponding to a sum of worldsheet
topologies. It remains an interesting open question whether such a
sum over topologies can be defined in the higher dimensional case.

\section{Asymptotic symmetry group}

To specify the asymptotic structure of a spacetime we attempt to construct
a set of boundary conditions that capture a wide-class of physically
interesting solutions. The Penrose conformal compactification of the
geometry provides an enormous simplification in treating these asymptotic
boundary conditions, because solutions may more easily be studied
on the compact unphysical spacetime (related by a Weyl transformation
to the physical spacetime) where the group of diffeomorphisms is clearly
defined \cite{Wald:1984rg}.

If one considers linearized perturbations around de Sitter, the conformal
group should have a well-defined action. In this limit, one can consider
the perturbation on top of the fixed de Sitter background, which has
as an isometry group $SO(d,1)$. These isometries induce a $SO(d,1)$
global conformal transformation on $\mathscr{I}$.

\subsection{Four dimensions}

At first sight, the situation for nonlinear solutions appears much
less clear. We will restrict our discussion to four-dimensional de
Sitter, and discuss the very special features of three dimensions
later. As mentioned in the introduction, already black holes will
tend to change the very topology of $\mathscr{I}$ and it is not clear
if any precise asymptotic conditions can be formulated. Ashtekar et
al. \cite{Ashtekar:2014zfa,Ashtekar:2015lla} deal with this by focussing
on isolated gravitating systems in de Sitter. Our approach will take
a different viewpoint, and allow for arbitrary boundary metrics that
respect the asymptotic de Sitter metric conditions locally
\begin{equation}
ds^{2}=R_{dS}^{2}\eta^{-2}\left(-d\eta^{2}+(\delta_{ij}+h_{ij})dx^{i}dx^{j}\right)\label{eq:metric}
\end{equation}
where we can perform a power series expansion of $h_{ij}$ as
\begin{equation}
h_{ij}(\eta,x)=h_{(0)ij}(x)+\eta^{2}h_{(2)ij}(x)+\eta^{3}h_{(3)ij}(x)+\mathcal{O}(\eta^{4})\label{eq:fmetric}
\end{equation}
following \cite{fefferman}. For now we will take $\mathscr{I}$ to
have topology of the 3-sphere, thus we are considering globally asymptotically
de Sitter spacetimes. We will consider more general topologies later
in the paper. As we will see later, typical matter configurations
only yield a single regular asymptotic region, so we take $\mathscr{I}$
to refer to either $\mathscr{I}^{+}$ or $\mathscr{I}^{-}$ but not
a disconnected union of the two.

One can largely separate the issue of topology change by first restricting
considerations to theories of gravity with conformally coupled matter.
There powerful nonlinear stability theorems have been proven by Friedrich
\cite{Friedrich:1991nn}. In particular, for an open set of initial
data, it has been shown that a past asymptotic de Sitter spacetime
can smoothly evolve to a future asymptotic de Sitter spacetime. The
solutions obtained involve a metric at $\mathscr{I}$ in a nontrivial
conformal class. These correspond to the usual long wavelength gravitons
of the theory of inflation, which freeze out when stretched past the
horizon scale. They induce a nontrivial Cotton tensor on $\mathscr{I}$.

As pointed out in \cite{Ashtekar:2014zfa,Ashtekar:2015lla}, demanding
conformal flatness of the boundary projects out these graviton modes
from de Sitter. Therefore if the holographic theory of de Sitter gravity
was simply a conformal field theory, living on a background with a
fixed conformal structure, the CFT would not be able to reproduce
the full set of graviton modes. One may of course perturbatively correct
for this by introducing sources on the boundary, however then one
must specify a path integral measure for such sources in order to
reproduce bulk observables, such as in-in correlators.

Let us try to establish the gauge symmetries of the boundary theory.
If we consider general asymptotic boundary conditions of the form
(\ref{eq:metric}) the asymptotic symmetry group is much larger than
the global conformal group. Instead, it consists of the full group
of diffeomorphisms of $\mathscr{I}$. As we will see later, we can
reconstruct part of the action of the holographic dual by considering
the boundary action of the bulk theory, evaluated on solutions of
the equation of motion. This boundary action then inherits the gauge
symmetry of the bulk, associated with diffeomorphisms of $\mathscr{I}$. 

The construction of the boundary theory is predicated on the Penrose
compactification of the bulk spacetime. This is achieved by performing
a general Weyl transformation of the bulk metric $g_{(unphys)\mu,\nu}=\Omega^{2}(\eta,x)g_{(phys)\mu,\nu}$
for some choice of smooth function $\Omega$ that vanishes on $\mathscr{I}$,
but with non-vanishing normal derivative. Again, by reconstructing
part of the action of the holographic dual involving the boundary
metric, one sees the boundary theory must inherit this Weyl invariance
as a gauge symmetry. We conclude then that the boundary theory must
be a theory of Euclidean conformal gravity.

In many ways, this is not a new statement. It has been advocated that
the dS/CFT correspondence be viewed as a computation of a wavefunction
via a CFT partition function
\begin{equation}
\Psi(h)=Z_{CFT}[h]\label{eq:wavefunction}
\end{equation}
where $h$ denotes the boundary metric. Our point is simply to compute
bulk observables, one must make the further step of computing
\[
\left\langle 0|O(x_{1})O(x_{2})|0\right\rangle =\int\mathcal{D}h\Psi^{*}(h)O_{CFT}(x_{1})O_{CFT}(x_{2})\Psi(h)=\int\mathcal{D}h\,Z_{CFT}^{*}[h]O_{CFT}(x_{1})O_{CFT}(x_{2})Z_{CFT}[h]
\]
with some a priori unknown measure $\mathcal{D}h$, and some de Sitter
spacetime operators $O$. Here the vacuum state $|0\rangle$ is to
be understood as an interacting generalization of the Bunch-Davies
vacuum. The operators $O_{CFT}$ are the dual CFT operators. For matter
fields in a fixed de Sitter background, these can be constructed \cite{Chatterjee:2015pha}.
To formulate a complete holographic description, one instead must
build the integration measure into the theory. This gives rise to
our conjecture that dS gravity is dual to a theory of conformal gravity
on $\mathscr{I}$. In that case, the relevant correlator would be

\begin{equation}
\left\langle 0|O(x_{1})O(x_{2})|0\right\rangle =\left\langle O_{cgrav}(x_{1})O_{cgrav}(x_{2})\right\rangle \label{eq:dsconformal}
\end{equation}
where the left-hand side is an in-in correlator in the bulk theory,
and the right-hand side represents the map of these observables into
the conformal gravity theory. The next goal is to try to specify as
much as possible, this conformal gravity theory. If this can be established,
it will then be necessary to revisit the boundary to bulk operator
mapping after properly understanding the gauge invariant observables
of the conformal gravity theory. In its current formulation \cite{Chatterjee:2015pha},
the mapping would only make sense for small perturbations around some
classical background.

\subsection{Quadratic action for holographic theory: 4d de Sitter}

In general to build operators in the boundary theory from those in
the bulk, one must use the integral transform method described in
\cite{Chatterjee:2015pha}, and its generalizations. This can be viewed
as an analog of the LSZ transform in constructing the S-matrix in
asymptotically flat spacetime.

In anti-de Sitter spacetime, one has a much easier task, because the
bulk to boundary mapping is much simpler, since the physical fields
of interest have simple power law falloff, dependent on their masses.
So while one must perform an integral transform to construct quasi-local
bulk fields from boundary operators, the inverse operation reduces
to taking a residue in the limit that the bulk operator approaches
infinity.

Nevertheless, if we focus on the gravitational field, and massless
minimally coupled scalars, for example, the results of AdS may be
continued to de Sitter. This is the approach followed in \cite{Maldacena:2002vr,Maldacena:2011mk}.
See also \cite{Harlow:2011ke} for related discussion of these issues.
Here let us generalize this to a massive scalar in de Sitter, with
action
\begin{equation}
S_{mat}=\int d\eta d^{3}x\,\frac{1}{2}\sqrt{-\mathrm{det}g}\left(-g^{\mu\nu}\partial_{\mu}\phi\partial_{\nu}\phi-m^{2}\phi^{2}-\xi R\phi^{2}\right)\,.\label{eq:scalaraction}
\end{equation}
In empty de Sitter, with metric (\ref{eq:metric}) the solution of
the equation of motion may be decomposed into the Bunch-Davies \cite{Chernikov:1968zm,Bunch}
mode functions
\[
u_{k}=\frac{1}{2^{5/2}\pi}\eta^{3/2}H_{\mu}^{(2)}(-k\eta)e^{ik\cdot x}
\]
where
\[
\mu^{2}=\frac{9}{4}-12\left(\frac{m^{2}}{R}+\xi\right)\,.
\]
Let us for the moment take $\mu$ to be real, corresponding to the
so-called complementary series representations of the conformal group.
Note we will work with the future half of the slicing, so $-\infty<\eta<0$.
We wish to compute the on-shell action, which reduces to a boundary
term as $\eta\to0^{-}$. We take a solution with some fixed behavior
on some late-time slice $\eta=\eta_{c}$
\[
\phi(\eta,x)=\frac{\eta^{3/2}H_{\mu}^{(2)}(-k\eta)}{\eta_{c}^{3/2}H_{\mu}^{(2)}(-k\eta_{c})}f_{\vec{k}}e^{i\vec{k}\cdot\vec{x}}
\]
and substitute into (\ref{eq:scalaraction}) to obtain
\begin{eqnarray}
iS_{mat} & = & iR_{dS}^{2}\int d^{3}x\frac{1}{2\eta^{2}}\phi\partial_{\eta}\phi|_{\eta=\eta_{c}}\nonumber \\
 & = & iR_{dS}^{2}\int\frac{d^{3}k}{2\left(2\pi\right)^{3}}f_{\vec{k}}f_{-\vec{k}}\frac{1}{4\eta_{c}^{3}}\left(3-2\mu-\frac{2\eta_{c}kH_{\mu-1}^{(2)}(-k\eta_{c})}{H_{\mu}^{(2)}(-k\eta_{c})}\right)\,.\label{eq:sccount}
\end{eqnarray}
When $\mu$ is half-integer, this expression may be expanded near
$\eta\to0^{-}$ (i.e. $\mathscr{I}^{+}$) and interpreted as a series
of counter-terms that must be subtracted to yield a finite boundary
action. For example, the massless minimally coupled scalar corresponds
to $m=0,\xi=0$ giving $\mu=3/2$ and
\[
iS_{mat}=R_{dS}^{2}\int\frac{d^{3}k}{2\left(2\pi\right)^{3}}f_{\vec{k}}f_{-\vec{k}}\left(\frac{ik^{2}}{2\eta_{c}}-\frac{k^{3}}{2}\right)
\]
as $\eta_{c}\to0$. The imaginary divergent term might then be subtracted
with a local $\int d^{3}x(\partial\phi)^{2}$ counterterm. The finite
piece yields the expected boundary propagator of a quasi-primary field
with conformal weight $\Delta=3=\frac{3}{2}+\mu$. The boundary action
for the scalar field then has the following form, which is non-analytic
in momenta
\[
S_{boundary}=R_{dS}^{2}\int d^{3}x\frac{1}{2}\phi\left(\square\right)^{3/2}\phi\,.
\]

Another simple example is the massless conformally coupled scalar,
with $m=0,\xi=1/6$ which gives $\mu=1/2$ and
\[
iS_{mat}=R_{dS}^{2}\int\frac{d^{3}k}{2\left(2\pi\right)^{3}}f_{\vec{k}}f_{-\vec{k}}\left(\frac{i}{2\eta_{c}^{3}}-\frac{k}{2\eta_{c}^{2}}\right)
\]
which has a vanishing finite boundary action after subtracting the
divergent counterterms. We will comment on this and the case of more
general mass in a moment.

In a transverse traceless gauge, the action for metric fluctuations
matches that of the massless minimally coupled scalar with a different
normalization, giving the boundary action
\[
S_{grav,boundary}=\frac{R_{dS}^{2}}{64\pi G}\int d^{3}x\,h_{ij}^{TT}\left(\square\right)^{3/2}h_{TT}^{ij}\,.
\]
As noted in \cite{Maldacena:2002vr} this gives a negative contribution
to the 2-point function of the boundary stress energy tensor proportional
to the central charge.

So far we have seen the boundary counter-term approach seems to work
well for the metric and massless minimally coupled scalar matter.
As noted in \cite{Chatterjee:2015pha} this approach of extracting
boundary operators for more general matter in de Sitter, by simply
taking asymptotic limits of the fields, fails in general. If one were
to evaluate (\ref{eq:sccount}) one would get oscillating cutoff (i.e.
$\eta_{c}$) dependent expressions\footnote{These oscillations play an important role in the minisuperspace approach
to solving the Wheeler-DeWitt equation \cite{Hartle:2008ng}. Our
philosophy in this section is to use the form of the square of this
wavefunction to guess the form of the conformal gravity action living
on $\mathscr{I}$. The duality proposed in (\ref{eq:dsconformal})
then makes no reference to the phase of this wavefunction.}. The correct approach is to apply an analog of the LSZ reduction
formula of asymptotically flat spacetime, by performing an integral
transform on the bulk fields to obtain a boundary expression that
transforms covariantly under the conformal group \cite{Chatterjee:2015pha}.
For scalar fields, this gives
\[
S_{matter,boundary}=R_{dS}^{2}\int d^{3}x\frac{1}{2}\phi\left(\square\right)^{\mu}\phi
\]
for $\mu>0$ real, corresponding to the complementary series of the
conformal group and 
\[
S_{matter,boundary}=R_{dS}^{2}\int d^{3}x\frac{1}{2}\left(\phi\left(\square\right)^{\mu}\phi+\phi^{*}\left(\square\right)^{-\mu}\phi^{*}\right)
\]
for $\mu$ imaginary, corresponding to the principal series of the
conformal group.

In the above, we have derived the quadratic terms that appear in the
conformal gravity theory coupled to matter. Already we see the conformal
gravity theory seems to be of a new kind, due to the non-polynomial
nature of its derivatives appearing in the quadratic term. The theory
appears to be free of ghosts, at least in the case when the matter
is restricted so $\mu>0$. At least at bulk tree-level, one should
be able to recover the higher order terms in the holographic Lagrangian.
The classical stability of de Sitter would seem to indicate this procedure
should be completely well-defined. 

An easy generalization of the above is to break parity in the bulk
by adding an $\int R\mathcircumflex R$ term, which corresponds to
adding a Chern-Simons gravity term to the boundary. This yields the
boundary Lagrangian for the well-studied case of topologically massive
gravity \cite{Deser:1981wh}, which is invariant under Weyl and diffeomorphism
symmetries.

As has been emphasized in \cite{Chatterjee:2015pha} the boundary
theory violates cluster decomposition, which is one of the central
axioms of Euclidean quantum field theory \cite{Osterwalder:1973dx,Osterwalder:1974tc}.
It will be very interesting to construct interacting holographic duals.
In the free limit, examples have been constructed in the context of
higher spin gravity in de Sitter \cite{Anninos:2011ui}.

\subsection{Three dimensions}

The case of three-dimensional de Sitter is special, because then $\mathscr{I}$
is two-dimensional, and always locally conformally flat. In this case,
the expansion of the metric (\ref{eq:metric}) takes the form 
\[
h_{ij}=h_{(0)ij}+\eta^{2}h_{(2)ij}+\mathcal{O}(\eta^{3})\,.
\]
We can nevertheless follow the strategy described above to compute
the boundary term arising from the on-shell bulk action. Now we will
find the boundary counter-term action
\[
iS_{grav}=\frac{i}{16\pi G_{3}}\int d^{2}x2\sqrt{-\mathrm{det}g}-\frac{1}{2}\log\left(-\eta_{c}\right)\sqrt{\mathrm{det}h_{(0)}}R_{(0)}\,.
\]
The anomalous contribution proportional to $\log\eta_{c}$ must be
cancelled for the theory to be conformally invariant. One way to approach
the problem is to couple the boundary theory to a Liouville field
theory with central charge adjusted so that a Weyl transformation,
shifting $\eta_{c}\to\alpha\eta_{c}$ is compensated by the anomaly
term coming from the Liouville theory. This renders the boundary theory
diffeomorphism invariant and Weyl invariant.

In this way, the boundary theory takes the form of the Polyakov string.
The central charge induced by the gravitational contribution to the
conformal anomaly is
\begin{equation}
c=-\frac{3R_{dS}}{2G_{3}}\,.\label{eq:ccharge}
\end{equation}
For the theory to be Weyl invariant at the quantum level, this central
charge must be cancelled by that of the Liouville field, leading to
a boundary theory with vanishing conformal anomaly.

In the usual conformal gauge of string theory, for fixed boundary
topology, the theory reduces to an ordinary conformal field theory
(coupled to the Liouville field) and the details of conformal gravity
may be forgotten. Moreover in string theory there is a well-defined
path integral involving sums over nontrivial worldsheet topologies.
Each topology is equipped with a well-defined moduli space. We expect
this sum over topologies is important to properly understand the holographic
theory describing quantum gravity in de Sitter, a question we turn
to in the next section.

\section{Topology change \label{sec:boundary topology}}

It is important for the consistency of AdS/CFT that there are strong
restrictions on the topology of the bulk geometry $M$ given the boundary.
For example, Witten and Yau \cite{Witten:1999xp} showed that the
boundary must be connected, and that the bulk Euclidean geometry satisfies
$H_{n}(M,\mathbb{Z})=0$ if the boundary has positive scalar curvature.

Similar topological restrictions have been explored in the context
of four-dimensional asymptotically de Sitter spacetimes in \cite{Andersson:2002nr,Galloway:2004bk}.
For example, if $\mathscr{I}^{+}$ has infinite fundamental group,
then one has the rather strong result if matter obeys the null energy
condition, there is no regular $\mathscr{I}^{-}$. Similarly if $\mathscr{I}^{+}$
has positive first Betti number, then the bulk is past null geodesically
incomplete. Nevertheless, there are many examples where at least $\mathscr{I}^{+}$
is well-defined. The case we will be most interested in is the case
where $\mathscr{I}^{+}$ is a sphere with punctures. Isolated gravitating
systems in de Sitter can reach $\mathscr{I}^{+}$ where they appear
as punctures. In the work of \cite{Ashtekar:2014zfa,Ashtekar:2015lla}
the focus is on a single isolated gravitating system. Since here we
are interested in building a holographic dual applicable to cosmology,
we will be most interested in is the case where $\mathscr{I}^{+}$
is a sphere with multiple punctures. 

If we wish to accommodate such isolated gravitating systems in the
dual conformal gravity theory living on $\mathscr{I}^{+}$, we must
therefore include a sum over topologies of the boundary. In the case
of three-dimensional Euclidean geometries, it is not clear whether
a path integral of conformal gravity over such a space can be defined.
Though it nevertheless appears to be a simpler problem than the original
proposals for four-dimensional Euclidean quantum gravity as a path
integral over geometries.

In the case of three-dimensional asymptotically de Sitter geometries
things are much simpler. Again, $\mathscr{I}^{+}$ is always conformally
flat, but one nevertheless must deal with this sum over topologies.
The sum over the moduli space of compact Riemann surfaces (including
punctures), is well-understood in the context of string theory and
leads to a complete proposal for the path integral of the conformal
gravity theory. That is, if we are given a Lagrangian for a CFT with
central charge (\ref{eq:ccharge}), we can couple it to conformal
gravity by performing a Weyl rescaling, and add in the Liouville sector
to cancel the overall conformal anomaly. One can then fix conformal
gauge, and treat the theory as one would with any worldsheet string
theory.

In the remainder of this section, we consider an example of a multi-black
hole solution in three-dimensional asymptotically de Sitter spacetime
\cite{Deser:1983nh}. If the above proposal is correct, it should
be possible to view $\mathscr{I}^{+}$ as a 2-sphere with punctures.
However the original work \cite{Deser:1983nh} expressed the Cauchy
slices as a multiple cover of a sphere with only two punctures at
the north and south poles. In the following, we construct the covering
space and show it is a single cover of a sphere with multiple non-degenerate
punctures.

\subsection{Example: multi-black hole solution in dS$_{3}$}

Deser and Jackiw have found the metric of 2+1 dimensional gravity
asymptotically de-Sitter spacetime \cite{Deser:1983nh} in the presence
of $N$ stationary massive particles. It is given by 
\begin{eqnarray}
ds^{2} & = & M^{2}(r)dt^{2}+f(r)dzdz^{*}\nonumber \\
f(z) & = & \frac{\epsilon}{\lambda V(z)V^{*}(z^{*})\cosh^{2}\left(\sqrt{\epsilon}\left(\zeta-\zeta_{0}\right)\right)}\nonumber \\
M(z) & = & \epsilon\tanh\left(\sqrt{\epsilon}\left(\zeta-\zeta_{0}\right)\right)\nonumber \\
V(z) & = & c^{-1}\prod_{n=1}^{N}(z-z_{n})\nonumber \\
\zeta(z) & = & \frac{1}{2}\left(\int\frac{dz}{V(z)}+\int\frac{dz^{*}}{V^{*}(z^{*})}\right)=\ln\left(\prod_{n}|z-z_{n}|^{c_{n}}\right)\nonumber \\
c_{n} & = & \prod_{n'\neq n}\frac{c}{z_{n}-z_{n'}}\nonumber \\
\sum_{n=1}^{N}c_{n} & = & 0\,.\label{eq:multiparticle ds}
\end{eqnarray}
Here $\lambda>0$ is the cosmological constant. The first equation
gives the metric in complex plane in terms of $f(z),M(z)$. $V(z)$
is the master function in terms of which the solution is given. $z_{n}$
are the punctures in the complex plane where particles are inserted
and $c$ is a free parameter. We demand that $c_{n}$ be real for
single valuedness of the solution. The coordinate transformation 
\begin{eqnarray}
\sin\omega & = & \frac{1}{\cosh\left(\sqrt{\epsilon}(\zeta-\zeta_{0})\right)}\nonumber \\
\phi & = & \frac{\epsilon}{2i}\left(\int\frac{dz}{V(z)}-\int\frac{dz^{*}}{V^{*}(z^{*})}\right)\label{eq:coord change to static}
\end{eqnarray}
takes us to the familiar static coordinates
\begin{eqnarray}
ds^{2} & = & -\cos^{2}\omega dt^{2}+\lambda^{-1}\left(d\omega^{2}+\sin^{2}\omega d\phi^{2}\right)\,.\label{eq:wphimetric}
\end{eqnarray}
Note that all the particles are located at $\sin\omega=0$ so that
$\omega=0,\pi$. The further coordinate change $\sqrt{\lambda}R=\sin\omega$
takes us to the static Schwarzschild-de-Sitter coordinates 
\begin{eqnarray*}
ds^{2} & = & -(1-\lambda R^{2})dt^{2}+(1-\lambda R^{2})^{-1}dR^{2}+R^{2}d\phi^{2}
\end{eqnarray*}
which covers the full space, but the range of $\phi$ goes from $[0,2\pi\alpha_{n})$
at the location of $n^{th}$ particle where $\alpha_{n}=\sqrt{\epsilon}c_{n}=1-4Gm_{n}$.
This is the familiar conical deficit of 3-dimensional gravity. Locally
the metric is same as pure de-Sitter and has constant curvature. 

Now we will investigate the geometry. First we will consider 3-particle
case before generalizing to the $N$ particle case. Uniqueness of
the solution requires that the 3 particles are all in a line and $c_{n}$
sum to zero. Let us take $c=1,z_{1}=-3,z_{2}=1,z_{3}=2$. Then let
us choose 
\begin{eqnarray*}
c_{1} & = & \frac{1}{\left(z_{1}-z_{2}\right)\left(z_{1}-z_{3}\right)}=\frac{1}{20}\\
c_{2} & = & -\frac{1}{4}\\
c_{3} & = & \frac{1}{5}\\
c_{1}+c_{2}+c_{2} & = & 0\,.
\end{eqnarray*}
Thus this configuration satisfies all the constraints. Now let us
look at the functions that determine the geometry
\begin{eqnarray*}
V(z) & = & (z+3)(z-1)(z-2)\\
\frac{1}{V(z)} & = & \frac{1}{20(z+3)}-\frac{1}{4(z-1)}+\frac{1}{5(z-2)}\\
\zeta & = & \frac{1}{2}\left(\int\frac{dz}{V(z)}+\int\frac{dz^{*}}{V^{*}(z^{*})}\right)=\frac{1}{20}\ln\left(\frac{|z+3||z-2|^{4}}{|z-1|^{5}}\right)\,.
\end{eqnarray*}
At $z=-3,2$, $\zeta=-\infty$ and at $z=1$, $\zeta=\infty$. In
$z$ coordinates, we have punctures at 3 points. We now want to understand
the picture in the $\omega,\phi$ coordinates using (\ref{eq:coord change to static})..

The points $z=-3,1,2$ correspond to $\sin\omega=0\implies\omega=0,\pi$.
Thus two of the particles are at south pole and one at north pole.
But then it is not immediately clear whether the particles at the
south pole are overlapping or they are multiple disconnected sheets
or they are sphere connected at some points etc. To understand the
topology, we first note that we can have a path between any two particles
without crossing the other particle. This implies that the sheets
are connected. Secondly, the distance between any two particles is
non-zero. 

\begin{figure}
\centering \includegraphics[scale=0.7]{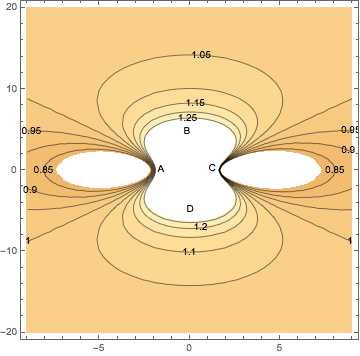}\label{contour}
\caption{Contour plot of $e^{20\zeta}=\frac{|z+3||z-2|^{4}}{|z-1|^{5}}$ in
the complex $z$-plane. }

\label{fig:contour} 
\end{figure}

To see this we note that constant $\omega$ corresponds to constant
$\frac{|z+3||z-2|^{4}}{|z-1|^{5}}$ curves in the complex plane. 
\begin{enumerate}
\item $\frac{|z+3||z-2|^{4}}{|z-1|^{5}}=\pm\infty$ would correspond to
north and south pole $\omega=0,\pi$.
\item $\frac{|z+3||z-2|^{4}}{|z-1|^{5}}=1$ corresponds to the equator $\omega=\pi/2$. 
\end{enumerate}
First we note that at $z=-3,2$, $\frac{|z+3||z-2|^{4}}{|z-1|^{5}}=0<1$
and at $z=1$, $\frac{|z+3||z-2|^{4}}{|z-1|^{5}}=\infty>1$. Thus
we are sure that $\frac{|z+3||z-2|^{4}}{|z-1|^{5}}=1$ contour will
pass between ($-3$ and $1$) and also between ($1$ and $2$). We
can verify it by plotting the contours as shown in Figure (\ref{fig:contour}).

The plot clearly shows that from $z=-3$ to $z=2$ we have to cross
$\frac{|z+3||z-2|^{4}}{|z-1|^{5}}=1$ contour at least twice. That
is we have to cross equator at least twice. To go from $z=-3,2$ to
$z=1$ we have to cross $\frac{|z+3||z-2|^{4}}{|z-1|^{5}}=1$ contour
or the equator at least once. Thus we are getting a picture where
we have two spheres. The south pole of one sphere corresponds to $z=-3$
and the south pole of other sphere correspond to $z=2$. The north
pole of both the spheres correspond to $z=1$. This means the two
spheres have common northern hemispheres ($\zeta>0$) and separate
southern hemispheres ($\zeta<0$).

\begin{figure}
\centering \includegraphics[scale=0.7]{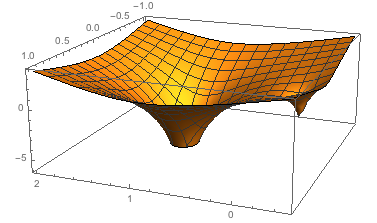}\label{pants} \caption{3D plot of $e^{20\zeta}$ after the transformation $z\rightarrow\frac{1}{z'}+1$
in z' plane. This shows that the topology of boundary is a pair of
pants. }

\label{fig:pants} 
\end{figure}

How does this all look in the $\omega,\phi$ coordinate? First let
us look at the contours for $\zeta>0\implies\omega<\pi/2$. These
curves are connected and close around $z=1.$ As we move along each
contour $\phi$ ranges from $[0,\alpha_{1}2\pi)$ where $1-\alpha_{1}$
is the conical deficit of the particle at $z=1$. For convenience
let us mark the point $A$ as $\phi=0$. Then as we move along the
curve we reach $B(\phi=\alpha_{1}\pi/2),C(\phi=\alpha_{1}\pi),D(\phi=3\alpha_{1}\pi/2)$
and when we come back to $A$, $\phi$ changes by $\alpha_{1}2\pi$.
These points are shown in figure (\ref{fig:contour}). This is true
for all the contours $\omega<\pi/2$. For $\omega=\pi/2$, contour
splits at $C,D\rightarrow\infty$. Topologically one then has a sphere
with 3 punctures, also know as the pants diagram. To see that explicitly,
we do the transformation $z\rightarrow\frac{1}{z'}+1$. This sends
$z=1,\infty$ to $z'=\infty,0$ respectively. The new function that
determines the geometry is shown in Figure (\ref{fig:pants}).

Generalizing to $N$ particles this will be $N$ punctures on the
Riemann sphere. Single valuedness of the solution requires that all
the punctures (position of the particles) lie on a line. Without loss
of generality we can take this line to be real axis. Mass defects
are given by absolute value of the $c_{n}$, which should sum to $0$.
We can choose $c=1$ in the solution (\ref{eq:multiparticle ds}).
and take the positions $z_{n}$ such that they satisfy $\sum c_{n}=0$.
Then the solution is given by (\ref{eq:multiparticle ds}). Let us
label the positions such that $z_{1}>z_{2}>...>z_{N-1}>z_{N}$. Then
\begin{eqnarray*}
c_{1} & = & \frac{1}{(z_{1}-z_{2})(z_{1}-z_{3})...(z_{1}-z_{N})}>0\\
c_{2} & = & \frac{1}{(z_{2}-z_{1})(z_{2}-z_{3})...(z_{2}-z_{N})}<0\\
c_{3} & > & 0\\
...
\end{eqnarray*}
$c_{2m+1}>0,c_{2m}<0$. Thus we see from equation (\ref{eq:multiparticle ds})
that $\zeta(z_{2m+1})=-\infty,\zeta(z_{2m})=\infty$. This solution
in $z$ coordinate is transformed to the de-Sitter like metric (\ref{eq:wphimetric})
using (\ref{eq:coord change to static}). We see that at $z=z_{n}\implies\omega=0,\pi$.
That is particles are either at the south or north pole, corresponding
to a multi-sheeted sphere with 2 punctures.

To see the geometry more clearly we look at the equator that is $\omega=\pi/2\implies\sin\omega=1\implies\zeta(z)=\zeta_{0}$
contours. Let us check these contours generate the Riemann sphere
with $N$ punctures. 

Since this is a compact manifold, all the contours must be closed.
The punctures live at $|\zeta(z_{n})|=\infty$. So none of the contours
with finite value of $\zeta_{0}$, end at the punctures. The second
observation is that the $\zeta(z)=0$ contour continuously extends
to $z=\infty$. Thus all the contours with $\zeta(z)=0$ are connected
at $z=\infty$. A third observation is that $\zeta(z_{2m+1})=-\infty,\zeta(z_{2m})=\infty$.
That is $\zeta(z)$ at successive punctures are of opposite sign.
Thus, $\zeta(z)=0$ contour separates any two successive punctures.
Thus there are $N-1$ $\zeta(z)=0$ contours joined at $z=\infty$.
These contours divide the Riemann sphere into $N$ segments. Each
segment contains exactly one puncture, and we have mapped the geometry
to a single-cover of the $N$-punctured sphere.

\section{Conclusion}

We have conjectured the holographic dual of an asymptotically de Sitter
spacetime in $d+1$-dimensions is a $d$-dimensional theory of Euclidean
conformal gravity living on $\mathscr{I}$. Various quadratic terms
in the action of the conformal gravity have been constructed, which
indicate the boundary metric becomes a dynamical variable. This then
forces one to consider whether the path integral over the boundary
metric includes a sum over topologies.

This is a sharp departure from the simplicity of the conformal field
theory/anti-de Sitter correspondence, where we have many examples
of suitable large $N$ conformal field theories and the boundary metric
is not dynamical. In the case of de Sitter, we instead get holographic
theories that violate the usual axioms of Euclidean field theory \cite{Chatterjee:2015pha}
and examples are hard to come by. The massless higher spin theories
have provided some examples where these issues can be explored in
detail \cite{Anninos:2011ui,Anninos:2013rza}. Optimistically one
might hope that the new feature of coupling to conformal gravity solves
some of these problems. More pessimistically it suggests that the
natural UV completion of de Sitter gravity may not be some lower dimensional
holographic theory, but is rather to be understood as an unstable
background in some larger complete theory \cite{Lowe:2007ek,Lowe:2010np}.

\paragraph*{Acknowledgements}

This research was supported in part by DOE grant DE-SC0010010 and
an FQXi grant. We thank Dionysios Anninos, Jennifer Traschen and David
Kastor for helpful discussions. We thank the U. of Montreal for hospitality
and the participants of the workshop on AdS/CFT, self-adjoint extensions
and the resolution of cosmological singularities for interesting discussions.

\bibliographystyle{utphys}
\bibliography{desittir}

\providecommand{\href}[2]{#2}\begingroup\raggedright\begin{thebibliography}{10}

\bibitem{Strominger:2001pn}
A.~Strominger, ``{The dS / CFT correspondence},''
  \href{http://dx.doi.org/10.1088/1126-6708/2001/10/034}{{\em JHEP} {\bfseries
  0110} (2001) 034},
\href{http://arxiv.org/abs/hep-th/0106113}{{\ttfamily arXiv:hep-th/0106113
  [hep-th]}}.

\bibitem{Chatterjee:2015pha}
A.~Chatterjee and D.~A. Lowe, ``{Holographic operator mapping in dS/CFT and
  cluster decomposition},''
\href{http://arxiv.org/abs/1503.07482}{{\ttfamily arXiv:1503.07482 [hep-th]}}.

\bibitem{Osterwalder:1973dx}
K.~Osterwalder and R.~Schrader, ``{Axioms for Euclidean Green's Functions},''
\href{http://dx.doi.org/10.1007/BF01645738}{{\em Commun.Math.Phys.} {\bfseries
  31} (1973) 83--112}.

\bibitem{Osterwalder:1974tc}
K.~Osterwalder and R.~Schrader, ``{Axioms for Euclidean Green's Functions.
  2.},''
\href{http://dx.doi.org/10.1007/BF01608978}{{\em Commun.Math.Phys.} {\bfseries
  42} (1975) 281}.

\bibitem{Abbott:1981ff}
L.~F. Abbott and S.~Deser, ``{Stability of Gravity with a Cosmological
  Constant},''
\href{http://dx.doi.org/10.1016/0550-3213(82)90049-9}{{\em Nucl. Phys.}
  {\bfseries B195} (1982) 76}.

\bibitem{Breitenlohner:1982jf}
P.~Breitenlohner and D.~Z. Freedman, ``{Stability in Gauged Extended
  Supergravity},''
\href{http://dx.doi.org/10.1016/0003-4916(82)90116-6}{{\em Annals Phys.}
  {\bfseries 144} (1982) 249}.

\bibitem{Ashtekar:2014zfa}
A.~Ashtekar, B.~Bonga, and A.~Kesavan, ``{Asymptotics with a positive
  cosmological constant: I. Basic framework},''
  \href{http://dx.doi.org/10.1088/0264-9381/32/2/025004}{{\em Class. Quant.
  Grav.} {\bfseries 32} no.~2, (2015) 025004},
\href{http://arxiv.org/abs/1409.3816}{{\ttfamily arXiv:1409.3816 [gr-qc]}}.

\bibitem{Ashtekar:2015lla}
A.~Ashtekar, B.~Bonga, and A.~Kesavan, ``{Asymptotics with a positive
  cosmological constant: II. Linear fields on de Sitter space-time},''
\href{http://arxiv.org/abs/1506.06152}{{\ttfamily arXiv:1506.06152 [gr-qc]}}.

\bibitem{Fradkin:1985am}
E.~S. Fradkin and A.~A. Tseytlin, ``{CONFORMAL SUPERGRAVITY},''
\href{http://dx.doi.org/10.1016/0370-1573(85)90138-3}{{\em Phys. Rept.}
  {\bfseries 119} (1985) 233--362}.

\bibitem{Horne:1988jf}
J.~H. Horne and E.~Witten, ``{Conformal Gravity in Three-dimensions as a Gauge
  Theory},''
\href{http://dx.doi.org/10.1103/PhysRevLett.62.501}{{\em Phys. Rev. Lett.}
  {\bfseries 62} (1989) 501--504}.

\bibitem{Berkovits:2004jj}
N.~Berkovits and E.~Witten, ``{Conformal supergravity in twistor-string
  theory},'' \href{http://dx.doi.org/10.1088/1126-6708/2004/08/009}{{\em JHEP}
  {\bfseries 08} (2004) 009},
\href{http://arxiv.org/abs/hep-th/0406051}{{\ttfamily arXiv:hep-th/0406051
  [hep-th]}}.

\bibitem{Witten:1999xp}
E.~Witten and S.-T. Yau, ``{Connectedness of the boundary in the AdS / CFT
  correspondence},'' {\em Adv. Theor. Math. Phys.} {\bfseries 3} (1999)
  1635--1655,
\href{http://arxiv.org/abs/hep-th/9910245}{{\ttfamily arXiv:hep-th/9910245
  [hep-th]}}.

\bibitem{Andersson:2002nr}
L.~Andersson and G.~J. Galloway, ``{DS / CFT and space-time topology},'' {\em
  Adv. Theor. Math. Phys.} {\bfseries 6} (2003) 307--327,
\href{http://arxiv.org/abs/hep-th/0202161}{{\ttfamily arXiv:hep-th/0202161
  [hep-th]}}.

\bibitem{Galloway:2004bk}
G.~J. Galloway, ``{Cosmological spacetimes with Lambda > 0},'' in {\em {The
  Beemfest: Advances in Differential Geometry and General Relativity Columbia,
  Missouri, May 10-11, 2003}}.
\newblock 2004.
\newblock \href{http://arxiv.org/abs/gr-qc/0407100}{{\ttfamily
  arXiv:gr-qc/0407100 [gr-qc]}}.
\newblock
[Submitted to: Contemp. Math.(2004)].

\bibitem{Deser:1983nh}
S.~Deser and R.~Jackiw, ``{Three-Dimensional Cosmological Gravity: Dynamics of
  Constant Curvature},''
\href{http://dx.doi.org/10.1016/0003-4916(84)90025-3}{{\em Annals Phys.}
  {\bfseries 153} (1984) 405--416}.

\bibitem{Wald:1984rg}
R.~M. Wald, {\em {General Relativity}}.
\newblock
1984.
\newblock

\bibitem{fefferman}
C.~Fefferman and C.~Graham, {\em Conformal invariants}, vol.~pp. 95-116.
\newblock Ast{\'e}risque (hors s{\'e}rie) Soci{\'e}t{\'e} Math{\'e}matique de
  France, Paris, 1985.

\bibitem{Friedrich:1991nn}
H.~Friedrich, ``{On the global existence and the asymptotic behavior of
  solutions to the Einstein-Maxwell-Yang-Mills equations},'' {\em J.Diff.Geom.}
  {\bfseries 34} (1991) 275--345.

\bibitem{Maldacena:2002vr}
J.~M. Maldacena, ``{Non-Gaussian features of primordial fluctuations in single
  field inflationary models},''
  \href{http://dx.doi.org/10.1088/1126-6708/2003/05/013}{{\em JHEP} {\bfseries
  05} (2003) 013},
\href{http://arxiv.org/abs/astro-ph/0210603}{{\ttfamily arXiv:astro-ph/0210603
  [astro-ph]}}.

\bibitem{Maldacena:2011mk}
J.~Maldacena, ``{Einstein Gravity from Conformal Gravity},''
\href{http://arxiv.org/abs/1105.5632}{{\ttfamily arXiv:1105.5632 [hep-th]}}.

\bibitem{Harlow:2011ke}
D.~Harlow and D.~Stanford, ``{Operator Dictionaries and Wave Functions in
  AdS/CFT and dS/CFT},''
\href{http://arxiv.org/abs/1104.2621}{{\ttfamily arXiv:1104.2621 [hep-th]}}.

\bibitem{Chernikov:1968zm}
N.~A. Chernikov and E.~A. Tagirov, ``{Quantum theory of scalar fields in de
  Sitter space-time},''
{\em Annales Poincare Phys. Theor.} {\bfseries A9} (1968) 109.

\bibitem{Bunch}
T.~S. {Bunch} and P.~C.~W. {Davies}, ``{Quantum field theory in de Sitter space
  - Renormalization by point-splitting},''
  \href{http://dx.doi.org/10.1098/rspa.1978.0060}{{\em Royal Society of London
  Proceedings Series A} {\bfseries 360} (Mar., 1978) 117--134}.

\bibitem{Note1}
These oscillations play an important role in the minisuperspace approach to
  solving the Wheeler-DeWitt equation \cite {Hartle:2008ng}. Our philosophy in
  this section is to use the form of the square of this wavefunction to guess
  the form of the conformal gravity action living on $\protect \mathscr {I}$.
  The duality proposed in (\ref {eq:dsconformal}) then makes no reference to
  the phase of this wavefunction.

\bibitem{Deser:1981wh}
S.~Deser, R.~Jackiw, and S.~Templeton, ``{Topologically Massive Gauge
  Theories},'' \href{http://dx.doi.org/10.1006/aphy.2000.6013,
  10.1016/0003-4916(82)90164-6}{{\em Annals Phys.} {\bfseries 140} (1982)
  372--411}.
[Annals Phys.281,409(2000)].

\bibitem{Anninos:2011ui}
D.~Anninos, T.~Hartman, and A.~Strominger, ``{Higher Spin Realization of the
  dS/CFT Correspondence},''
\href{http://arxiv.org/abs/1108.5735}{{\ttfamily arXiv:1108.5735 [hep-th]}}.

\bibitem{Anninos:2013rza}
D.~Anninos, F.~Denef, G.~Konstantinidis, and E.~Shaghoulian, ``{Higher Spin de
  Sitter Holography from Functional Determinants},''
  \href{http://dx.doi.org/10.1007/JHEP02(2014)007}{{\em JHEP} {\bfseries 1402}
  (2014) 007},
\href{http://arxiv.org/abs/1305.6321}{{\ttfamily arXiv:1305.6321 [hep-th]}}.

\bibitem{Lowe:2007ek}
D.~A. Lowe, ``{Some comments on embedding inflation in the AdS/CFT
  correspondence},'' \href{http://dx.doi.org/10.1103/PhysRevD.77.066003}{{\em
  Phys.Rev.} {\bfseries D77} (2008) 066003},
\href{http://arxiv.org/abs/0710.3564}{{\ttfamily arXiv:0710.3564 [hep-th]}}.

\bibitem{Lowe:2010np}
D.~A. Lowe and S.~Roy, ``{Punctuated eternal inflation via AdS/CFT},''
  \href{http://dx.doi.org/10.1103/PhysRevD.82.063508}{{\em Phys.Rev.}
  {\bfseries D82} (2010) 063508},
\href{http://arxiv.org/abs/1004.1402}{{\ttfamily arXiv:1004.1402 [hep-th]}}.

\end{thebibliography}\endgroup

\end{document}